# Temperature effect observed by the Nagoya muon telescope

M. Berkova, A. Abunin, M. Preobrazhensky, S. Karimov, V. Yanke
*IZMIRAN, Moscow, RU-142190, Russia*

The multidirectional muon telescope at Nagoya (35°09'N, 136°58'E) is the most successful at the point of construction multidirectional scintillation telescope. It is working since 1970 and has 17 independent directions: a vertical, on 4 inclined 30º, 49º and 64º and 4 azimuthal directions. The temperature coefficients for all the directions of the Nagoya muon telescope were obtained using three different methods for the temperature effect calculation. The zenith angular dependence of the temperature coefficients was studied. Also, using the long–term data (from 1986 to 2016) of the Nagoya telescope a set of the yearly temperature coefficients was obtained and analyzed.

## 1. INTRODUCTION

This work continues a series of publications [1-7] on the study of the temperature effect of the muon cosmic ray component. In these works the main attention was paid to the development of the methods of eliminating the temperature effect from the data of muon detectors of different geometries and locations (mountain, ground and underground) with the help of altitude temperature profile of the Global Forecast System (GFS) model [8]. The purpose of this work is to study the long-term changes in the temperature coefficients and their dependence on zenith angles, i.e. on the effective energy of the detected particles, according to the Nagoya muon telescope data [9, 10].

## 2. METHODS OF THE TEMPERATURE EFFECT EXCLUSION

In general there are three main methods of temperature effect calculation. The integral method is presented in [11-13]. Temperature variations are determined as

$$\delta I_\mu / I_{\mu_{Temp}} = \int_0^{h_0} \alpha(h) \cdot \delta T(h) dh \quad (1)$$

where $\delta I_\mu / I_{\mu_{Temp}}$ is temperature variations of muon intensity, $\delta T(h)$ is temperature variations from the base period $\delta T(h) = T_{Base}(h) - T(h)$. So–called differential temperature coefficients $\alpha(h)$ have a dimension of [%/°K·atm], and they are theoretically calculated for different depths and angles. Usually the effective temperature method [14] is used, and it is simply another form of the integral method. Indeed, normalizing and simultaneously multiplying (1) by the temperature coefficient

$$\alpha_T = \int_0^{h_0} \alpha(h) dh \quad [\%/°K] \quad (2)$$

we obtain:

$$\left. \frac{\delta I_\mu}{I_\mu} \right|_{Temp} = \int_0^{h_0} \alpha(h) dh \cdot \frac{\int_0^{h_0} \alpha(h) \delta T(h) dh}{\int_0^{h_0} \alpha(h) dh} = \alpha_T \cdot \delta T_{eff} \quad (3)$$

where the effective temperature is:

$$T_{eff} = \frac{1}{\int_0^{h_0} \alpha(h) dh} \int_0^{h_0} \alpha(h) \delta T(h) dh \quad [°K] \quad (4)$$

There is a method of mass–average temperature. It was first noted in [15]. As the densities of temperature coefficient $\alpha(h)$ for the ground–based detectors are not strongly changed with the atmospheric depth $h$, the average $\overline{\alpha(h)}$ can be put out behind the integral sign (2) as follows:

$$\delta I_T = \overline{\alpha}_T \int_0^{h_0} \delta T(h) \cdot dh = \overline{\alpha}_T \sum_{n=1}^{L_{skin}} \frac{\Delta h_n}{h_0} \cdot \overline{T}_n = \overline{\alpha}_T \cdot \delta T_m \quad (5)$$

where $T_m$ is mass–average temperature and $L_{skin}$ is a surface layer number.

In addition there is an alternative empirical method. The method of effective level of generation is based on the assumption that muons are generally generated at the isobaric level usually taking for 100 mb, and its height is varying with change of the atmosphere temperature. According to [16, 17] the variation of the muon intensity is correlated with the height of generation level $\delta H$ and with the air temperature in this layer $\delta T$, i.e.

$$\delta I_T = \alpha_H \delta H + \alpha_T \delta T \quad (6)$$

where $\alpha_H$ (%/km) is so–called decay factor – the negative effect, and $\alpha_T$ (%/C$^0$) – the positive temperature coefficient.

## 3. TEMPERATURE DATA

In this work the temperature model data of the Global Forecast System (GFS) representing by the NCEP (National Centers for Environmental Prediction – USA) has been used [8, 18]. The GFS model makes it possible to obtain both retrospective and prognostic data of a 3D temperature field. The model's outputs are temperature data at 17 standard isobaric levels four times a day (00, 06, 12, 18UT). To obtain hourly data, interpolation by the cubic spline function on nine nodal points is carried out. The original data of the muon supertelescope (archived and real-time) can be found at [9, 10].

## 4. TEMPERATURE COEFFICIENT VS ZENITH ANGLE

In the work [3], the temperature effect for the Nagoya telescope was determined by three methods described above, but only for the vertical direction. The aim of this study was to determine the temperature effect for all 17 independent directions of the Nagoya telescope. To obtain the temperature coefficients, the data for 2009 were used, as it was a year of minimum solar activity. Therefore, the observed variations can be explained mainly by the meteorological effects. Thus, the regression coefficients for all 17 directions of the Nagoya telescope for 2009 were obtained by the methods described above, and they are presented in Table 1.

Table 1: Regression coefficients for all the directions of the Nagoya muon telescope for 2009

| Compo-nent | Effective temperature method | | Mass–average method | | Duperier method | |
|---|---|---|---|---|---|---|
| | $\alpha_{Teff}$ %/C° ± 0.015 | $\rho$ ± 0.01 | $\alpha_{Tm}$ %/C° ± 0.016 | $\rho$ ± 0.01 | $\alpha_{H100}$ %/km | $\rho$ ± 0.01 |
| Vertical | -0.239 | -0.97 | -0.231 | -0.97 | -4.363 | -0.94 |
| 30° N | -0.251 | -0.97 | -0.241 | -0.97 | -4.560 | -0.94 |
| 30° S | -0.253 | -0.97 | -0.243 | -0.97 | -4.592 | -0.94 |
| 30° E | -0.249 | -0.97 | -0.239 | -0.97 | -4.522 | -0.95 |
| 30° W | -0.252 | -0.96 | -0.242 | -0.97 | -4.566 | -0.94 |
| 39° NE | -0.258 | -0.97 | -0.245 | -0.97 | -4.659 | -0.95 |
| 39° NW | -0.262 | -0.96 | -0.249 | -0.96 | -4.722 | -0.94 |
| 39° SE | -0.259 | -0.97 | -0.246 | -0.97 | -4.682 | -0.94 |
| 39° SW | -0.262 | -0.96 | -0.249 | -0.96 | -4.725 | -0.94 |
| 49° N | -0.274 | -0.96 | -0.255 | -0.96 | -4.857 | -0.94 |
| 49° S | -0.271 | -0.96 | -0.252 | -0.96 | -4.806 | -0.94 |
| 49° E | -0.269 | -0.97 | -0.251 | -0.97 | -4.782 | -0.95 |
| 49° W | -0.272 | -0.96 | -0.253 | -0.96 | -4.821 | -0.94 |
| 64° N | -0.269 | -0.93 | -0.237 | -0.94 | -4.513 | -0.92 |
| 64° S | -0.269 | -0.93 | -0.236 | -0.93 | -4.509 | -0.91 |
| 64° E | -0.259 | -0.93 | -0.227 | -0.94 | -4.338 | -0.92 |
| 64° W | -0.267 | -0.93 | -0.235 | -0.93 | -4.474 | -0.91 |

The results show that the temperature coefficients, obtained by means of the effective and the mass-average temperature methods, are differ within <1%. Apparently, a little large (in amplitude) values of the temperature coefficients, obtained by the effective temperature method, can be explained by the higher layers of the atmosphere. In the calculation of the effective temperature, the contribution of the high atmospheric layers are counted with larger weights than the lower ones. The accuracy of the results confirms by the strong correlation coefficients for all the directions and for all the methods.

Lower correlation coefficients were obtained for directions with the largest zenith angle 64° due to the significantly lower intensity of muons. And in general, there is less correlation for the method of effective level of generation (the Duperie method). This can be explained by significantly large errors in measuring the height of 100 mb level.

As an example, in Figure 1 there are plots of the effective temperature (blue curve), the vertical count rate variations, uncorrected for the temperature effect (red curve), and the vertical count rate variations, corrected for the temperature effect by the effective temperature method (yellow curve), by the mass–average temperature method (green curve), by the Duperier method (black curve). It is seen that after the data were corrected for the temperature effect, the annual temperature wave disappeared. At the same time there is good agreement between the two methods of effective and mass-average temperatures - almost merged yellow and green curves (Figure 1a). The Duperie method showed a much worse result. From good agreement of variations, corrected for the temperature effect by the effective temperature method and by the mass-average temperature method, we can conclude that the theoretical differential temperature coefficients $\alpha(h)$, that are used in this work, are correct.

## 5. TEMPERATURE COEFFICIENTS: EXPERIMENT AND THEORY

An analysis of the angular dependence of the temperature coefficients were held for a calm 2009. For this purpose the temperature coefficients were grouped by azimuthal directions (N-E-W-S). In Figure 2 the values of 17 temperature coefficients, obtained by three methods for 2009, are shown - temperature coefficients, calculated by the mass-average method $\alpha_M$ (squares, the left axis), by the effective temperature method $\alpha E$ (circles, the left axis) and by the Duperier method $\alpha H$ (circles, the right axis). At the same Figure 2 there is also a theoretical curve of the temperature coefficients dependence on zenith angle (C curve), calculated in accordance with the expression (2), in which the differential temperature coefficients $\alpha(h)$ are taken from the work [19].

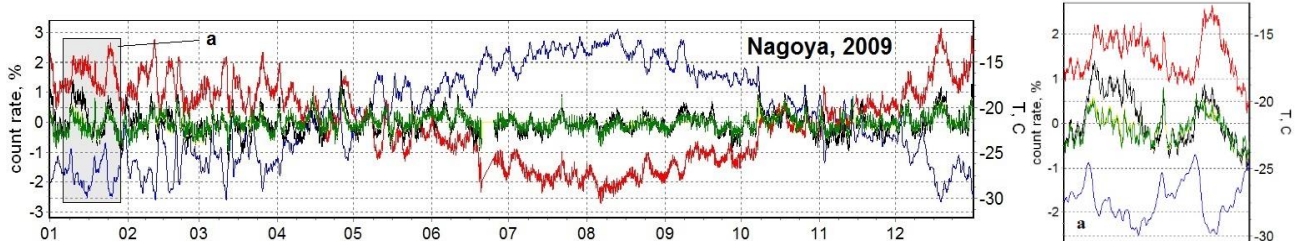

Figure 1: The Nagoya muon telescope, 2009. Uncorrected for the temperature count rate variations for the vertical – red curve, the effective temperature – blue curve. Corrected for the temperature count rate variations: by the effective temperature method – yellow curve, by the mass–average temperature method – green curve, by the method of effective level of generation – black curve.

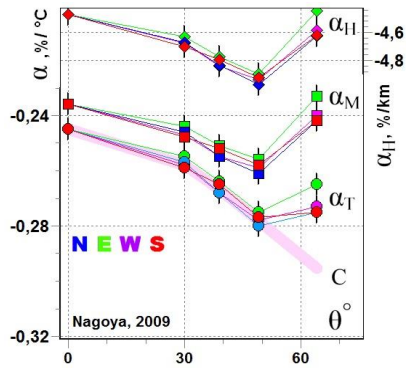

Figure 2: The angular dependence of the temperature coefficients for the Nagoya muon telescope

It is seen in Figure 2 that independently of the calculation method, the negative temperature coefficients grow in amplitude up to about 50 ° with the model values. At large zenith angles the temperature coefficients start to deviate from the model curve towards reduction of the absolute values. Most likely, this can be explained by decrease in the negative and by increase in the positive component of the temperature effect and, as a result, decrease in the total negative temperature coefficient

It should be noted that the methods of effective and mass-average temperatures and the Duperie method are fundamentally different methods. The first two are based on consideration of the nuclear cascade process in the atmosphere, and Duperie method – an empirical method and does not require any model assumptions. However, the angular dependence of the experimentally obtained temperature coefficients behaves in all three cases exactly in the same way - an anomalous behavior at high angles.

The total negative temperature effect of the muon component, recorded by the ground-based telescopes, determined by short-lived low-energy muons, sensitive to the path length they have to pass before decay. With increasing inclination angle the geometric path of muons is increased, and that leads to a cutoff of the lowest-energy muons, as they decay before reach the ground level. The median regidity for Nagoya muon telescope is in the range of 50 GV (vertical) to 110 GV ($\theta=64°$) [20]. And if muons with energies of a few tens GeV are more sensitive to their path length, then the muons of ~ 100 GeV are much less sensitive to the changes of the atmosphere height.

Along with the increase of zenith angle the parent particles of muons have to travel relatively long distances in rare parts of the atmosphere. As a consequence, their decay probability is increased compared to probability of their interaction with the atmosphere nuclei. This effect leads to the positive temperature component increase.

The anomalous behavior of the regression coefficients, obtained by the Duperier method, at high angles may be explained in the same way - muons of ~ 100 GeV are much less sensitive to the fluctuations of the 100mb level height caused by the temperature changes.

It is worth noting that you can do not take into account the curvature of the Earth for the considered range of zenith angles. Perhaps, deviation of the experimentally found temperature coefficient is attributed to singularities of the angular distribution of the muons in the atmosphere and its improper accounting.

## 6. TEMPERATURE EFFECT: 1980-2016

To analyze the long-term range of the temperature coefficients, using the data from 1980 to 2016, the temperature coefficients for the vertical were determined by the effective temperature method. The results are shown in the Table 2.

Table 2: Temperature coefficients, calculated by the effective temperature method for the Nagoya muon telescope (vertical) for 1980-2016

| Year | $\alpha_{Teff}$, %/C° ± 0.022 | ρ ± 0.014 | Year | $\alpha_{Teff}$, %/C° ± 0.022 | ρ ± 0.014 |
|---|---|---|---|---|---|
| 1980 | -0.221 | -0.918 | 1999 | -0.260 | -0.953 |
| 1981 | -0.235 | -0.927 | 2000 | -0.390 | -0.852 |
| 1982 | -0.303 | -0.895 | 2001 | -0.258 | -0.955 |
| 1983 | -0.196 | -0.889 | 2002 | -0.273 | -0.941 |
| 1984 | -0.236 | -0.929 | 2003 | -0.287 | -0.946 |
| 1985 | -0.218 | -0.789 | 2004 | -0.266 | -0.959 |
| 1986 | -0.202 | -0.804 | 2005 | -0.287 | -0.962 |
| 1987 | -0.272 | -0.885 | 2006 | -0.279 | -0.957 |
| 1988 | -0.226 | -0.930 | 2007 | -0.257 | -0.964 |
| 1989 | -0.215 | -0.864 | 2008 | -0.256 | -0.979 |
| 1990 | -0.211 | -0.909 | 2009 | -0.245 | -0.977 |
| 1991 | -0.269 | -0.901 | 2010 | -0.267 | -0.979 |
| 1992 | -0.205 | -0.896 | 2011 | -0.268 | -0.976 |
| 1993 | -0.232 | -0.949 | 2012 | -0.255 | -0.964 |
| 1994 | -0.244 | -0.924 | 2013 | -0.252 | -0.963 |
| 1995 | -0.242 | -0.946 | 2014 | -0.229 | -0.951 |
| 1996 | -0.273 | -0.933 | 2015 | -0.226 | -0.962 |
| 1997 | -0.244 | -0.965 | 2016 | -0.224 | -0.963 |
| 1998 | -0.258 | -0.942 | | | |

For 37 years, the temperature coefficient varies within ~ (-0.24 ± 0.03)% / C ° (except for a few years). In the years of minimum solar activity the correlation is highest since the weather conditions are mainly responsible for the observed variations - minimally pressure, mostly - the atmospheric temperature.

## 7. CONCLUSIONS

A set of regression temperature coefficients for the Nagoya muon telescope was get by three methods: the effective temperature method, the mass–average temperature method and the method of effective level of generation. For the ground–based telescopes correction for the temperature effect, obtained by the mass–average temperature method, is a very good approximation and is almost identical to the results obtained by the method of the effective temperature, despite of the fact that the atmosphere layers are taken into account with equal weights. The method of effective level of generation is the least accurate of all the methods.

A long–term set of temperature coefficients for the Nagoya telescope (vertical), calculated by the effective temperature method for the period of observation 1980–2016 was obtained. The average value of the temperature coefficient for the vertical is (–0.24±0.03) [%/°C].

For 2009 the temperature coefficients for all directions and zenith angles of particle arrival were calculated for the Nagoya muon telescope. The temperature coefficients for zenith angles >50° are significantly diverged from the model curve towards decreasing the absolute values. This may mean the negative component decrease and increase of the positive component of the temperature effect.

The data of the Nagoya muon telescope over the entire observation period can be found in the database mddb [21], which united the original data, the data corrected for the temperature effect and the related meteorological data: pressure, temperature distribution on the standard isobaric levels. All data have hourly time resolution.

## Acknowledgments

This work was partially supported by the Program of the Presidium of RAS № 23 "high energy physics and neutrino astrophysics," RFBR grant № 17-02-00508, experimentally and methodologically support the project UNU №85 "Russian national network of ground stations of cosmic rays." We are grateful to all the staff of the World Network of CR stations http://cr0.izmiran.ru/ThankYou